# Star Formation Around Supermassive Black Holes

I.A. Bonnell[1], W.K.M. Rice[2]

[1]*Scottish Universities Physics Alliance, University of St Andrews, Physics & Astronomy, North Haugh, St Andrews, Fife KY16 9SS, UK.*

[2]*Scottish Universities Physics Alliance, Institute for Astronomy, University of Edinburgh, Blackford Hill, Edinburgh EH9 3HJ, UK.*

**The presence of young massive stars orbiting on eccentric rings within a few tenths of a parsec of the supermassive black hole in the Galactic centre is challenging for theories of star formation. The high tidal shear from the black hole should tear apart the molecular clouds that form stars elsewhere in the Galaxy, while transporting the stars to the Galactic centre also appears unlikely during their stellar lifetimes. We present numerical simulations of the infall of a giant molecular cloud that interacts with the black hole. The transfer of energy during closest approach allows part of the cloud to become bound to the black hole, forming an eccentric disc that quickly fragments to form stars. Compressional heating due to the black hole raises the temperature of the gas to 100-1000K, ensuring that the fragmentation produces relatively high stellar masses. These stars retain the eccentricity of the disc and, for a sufficiently massive initial cloud, produce an extremely top-heavy distribution of stellar masses. This potentially repetitive process can therefore explain the presence of multiple eccentric rings of young stars in the presence of a supermassive black hole.**

By tracking the motions of young massive stars, two teams of astronomers have uncovered the existence of a $3.6 \times 10^6$ M$_o$ (solar masses) supermassive black hole (SMBH) in the center of our Galaxy (1-3). In addition, one, possibly two, eccentric rings of young massive stars orbit slightly further out, near ~0.1 pc (4), while low-mass stars are sparse in the region (5), making our Galactic center the best example of an atypical distribution of stellar masses (or stellar initial mass function, IMF).

The presence of young massive stars in the vicinity of the Galactic centre is difficult to reconcile with current models of star formation where turbulent molecular clouds produce a mostly clustered population of stars with a remarkably constant distribution of stellar masses, covering stars with masses from less than a tenth to greater than 100 times the mass of the sun (6). The tidal pull from the supermassive back hole should disrupt anything as tenuous as a molecular cloud, thereby destroying it before it can even form and thus removing the necessary conditions for star formation (7). This leaves the possibilities that either the stars formed elsewhere and migrated to the Galactic center, or that the stars formed in situ by an unusual mechanism, such as the fragmentation of a gaseous disc rotating around the supermassive black hole (8,9).

A stellar cluster could migrate to the Galactic center by losing energy through its gravitational disturbance of the background stars in the Galaxy. However, this appears to take too long to explain the existence of the young stars found there (10). Star formation in an accretion disk around the supermassive black hole is possible if the disc is sufficiently massive (11), and if it is able to cool sufficiently rapidly (12,13). A priori a disc should fragment producing circular orbits, not the eccentric rings recently detected. Dynamical relaxation of a top-heavy IMF could increase the eccentricities of individual stars but is unlikely to reach the values observed suggesting that the stars are likely to have formed from an initially eccentric disc (14,15). Here, we explore how the accretion disc forms around the black hole, and in particular how it could form with an initial eccentricity, by simulating the evolution of a giant molecular cloud (GMC) falling towards a supermassive black hole (16). The cloud is presumed to be the result of a collision at a distance of several pc from the Galactic centre, sending it on this plunging orbit towards the black hole.

We carried out the simulations using Smooth Particle Hydrodynamics (SPH), a Lagrangian hydrodynamics formalism (17). We considered two situations, a $10^4$ $M_o$ molecular cloud falling towards a $10^6$ $M_o$ black hole, and a $10^5$ $M_o$ molecular cloud falling towards a $3 \times 10^6$ $M_o$ black hole. The clouds were initially placed 3 pc from the black hole and on an orbit with an impact parameter of ~ 0.1 pc. The clouds were turbulently supported and had a minimum temperature of 100 K due to the background radiation field. We included an approximate radiative transfer formalism with compressional heating balanced by cooling rates derived from estimated optical depths (18). Dense protostellar fragments were replaced by sink-particles (19) that accreted all bound gas particles that fell within 200 AU. Additional information as to the details of the simulation are available in the supplementary online material. Our computations were performed using the U.K. Astrophysical Fluids Facility (UKAFF) and our local SUPA HPC facility.

The early evolution of both clouds was broadly similar over the ~ 2 x $10^4$ years required to reach the Galactic center. As the $10^4$ $M_o$ molecular cloud fell towards the $10^6$ $M_o$ black hole (Fig. 1), it became tidally distorted, and unbound, due to the strong gravitational field. The turbulence in the cloud formed local structures some of which collapsed to form stars before the cloud reached the black hole. Stars formed before closest approach remained unbound and escaped the system. In contrast, 10 per cent of the gas cloud became bound due to the combination of gas dissipation and tidal torques at closest approach and remained in orbit within a few tenths of a pc. The tidal disruption of the self-gravitating cloud formed coherent spiral structures that transferred orbital angular momentum outwards while shocks from infalling streams, which passed either side of the black hole, also removed orbital angular momentum from the gas, allowing it to form an eccentric disc-like structure.

Structure in the infalling gas ensured that the disk of bound gas was very clumpy, forming spiral patterns that grew due to the disk's self-gravity and fragmented to form individual stars. The fragmentation occurred very quickly (on an orbital timescale) to form 498 stars with eccentricities between $e = 0.6$ and $e = 0.76$ and semi-major axes between $a = 0.11$ pc and $a = 0.19$ pc. The larger $10^5$ M$_o$ cloud (the final state is shown in Fig. 2) formed 198 stars with semi-major axes between $a = 0.02$ pc and $a = 0.13$ pc and eccentricities between $e = 0$ and $e = 0.53$. The larger mass and size of the cloud resulted in more angular momentum being removed by the tidal torques and by direct shocks on the incoming streams such that the eccentric disc and stars were formed closer in to the black hole. The stellar discs were fairly thin with an initial H/R varying from 0.1 to 0.2. Dynamical relaxation of these disks would increase the H/R slightly over several million years.

The resulting masses of the stars formed depends crucially on the balance between compressional heating of the gas as it infalls towards the black hole and the radiative cooling given by the gas temperature and the local optical depth. Near the black hole, the heating dominated, increasing the gas temperature to between several hundred and several thousand degrees K. This resulted in a Jeans mass, the minimum fragment mass, of order 0.5 $M_o$ for the $10^4$ $M_o$ cloud and up to 10 to 50 $M_o$ for the $10^5$ $M_o$ cloud. The higher mass cloud produced higher temperatures and hence higher Jeans masses as more gas was captured closer in to the black hole. This increased the compressional heating and decreased the cooling rate due to the higher the optical depths. Stars that formed further out, near 0.1 pc, had masses of several $M_o$ as in the lower-mass cloud.

The stellar mass functions from the two simulations (Fig. 3 and Fig. 4) are very different: the lower-mass cloud formed a typical stellar mass distribution peaking at 0.8 $M_o$ and following a Salpeter-type power law at higher-masses [$dN(\log m) = m^{-\Gamma} d(\log m)$, with $\Gamma \sim 1.35$]. The higher-mass cloud produced a bimodal mass function: a population of very massive stars with masses between ~ 10 and ~100 $M_o$, and a population of lower mass stars. The higher-mass stars formed in the inner ring (a~0.02 pc) while the lower-mass stars formed further out (a~ 0.05-0.1 pc) due to the different gas temperatures produced (see online material). As additional gas remained bound at larger radii, it is possible that more lower-mass stars would eventually form if the simulation was followed further in time.

In addition to forming the stars, 10 to 30 % of the infalling gas clouds were accreted onto the black hole. This accretion implies only that the material is bound within the size of the sink-particle's accretion radius of 4000 AU, and in fact this material had sufficient angular momentum to form a disk at radii of 1000-4000 AU around the black hole.

The actual size and evolution of this inner disk is not determined by our simulations and could involve further star formation or self-gravity driven accretion. Assuming that all the gas would accrete directly onto the black hole on a viscous timescale of order $10^7$ years yields an average mass accretion rate of order $10^{-4}$ $M_o$ / yr for the $10^4$ $M_o$ cloud and $10^{-3}$ $M_o$ / yr for the $10^5$ $M_o$ cloud. This gives a maximum accretion luminosity of order $10^{43}$ ergs/s, or about 1 % of the Eddingtom luminosity. This additional source of radiation, and that from the newly formed stars could increase the gas temperature in the disk and thus the fragment masses.

Our simulations show that an infalling molecular cloud can indeed form an eccentric disc around a supermassive black hole and that although the tidal force of the black hole will disrupt the cloud, it does not destroy the small-scale structures that seed the disk fragmentation. Furthermore, the compressional heating of the infalling gas results in the formation of a population of stars biased towards higher masses. The stellar masses depend crucially on the mass of the infalling cloud and on its impact parameter allowing for a variety of final outcomes. The initial disc eccentricity also means that the stars can form with initial eccentricities and, if the molecular cloud is sufficiently massive, the stars that form may be extremely massive. This is, therefore, a viable mechanism for forming the rings of young, massive stars within ~ 0.1 pc of the galactic centre. What is still unclear however, is the origin of the infalling cloud and the probability of the small impact parameter that is required.

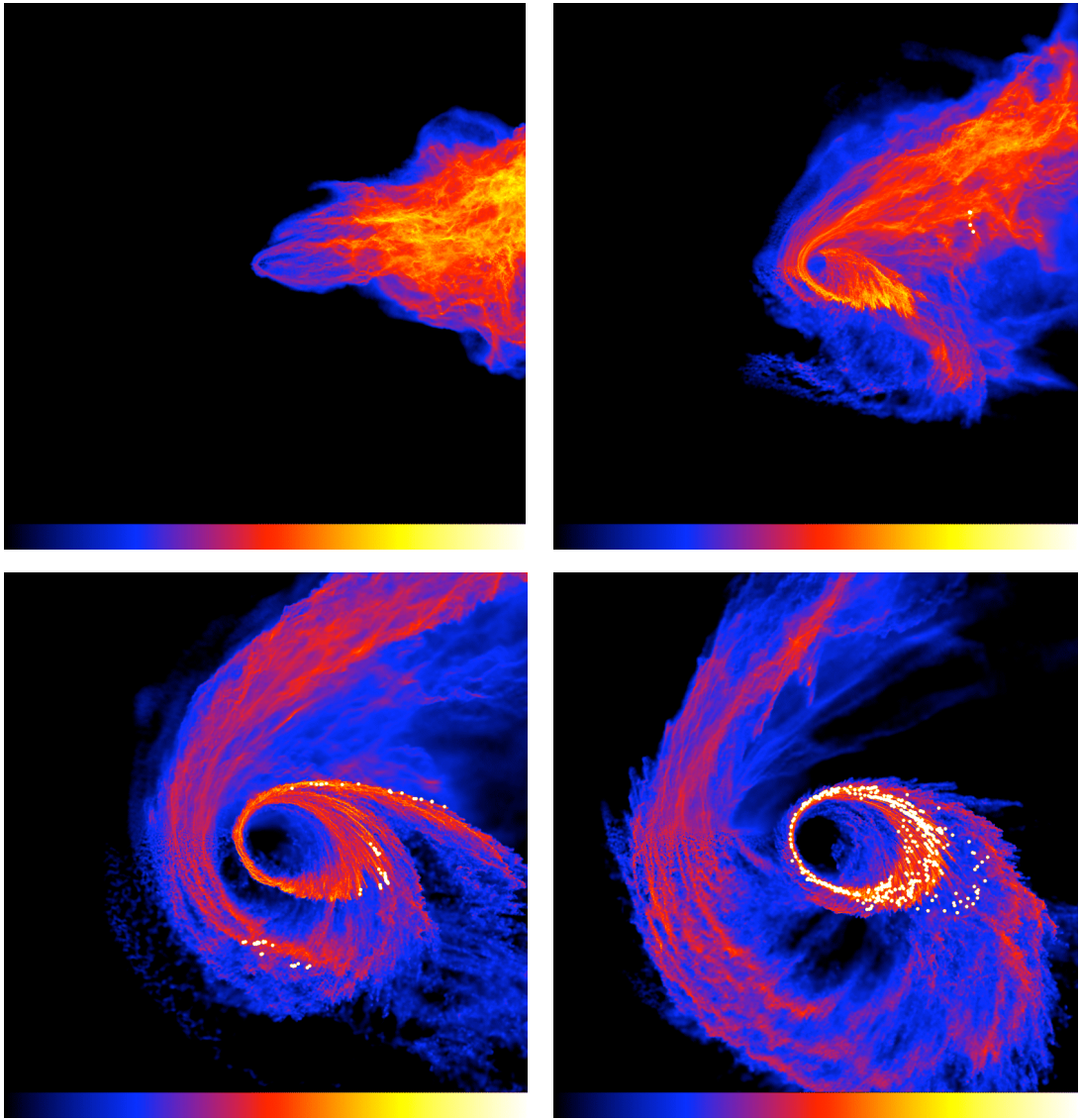

**Fig 1.** The evolution of a $10^4$ M molecular cloud is shown falling toward a $10^6$ M black hole. The upper left image shows the region within 1.5 pc of the black hole and the colors illustrate the column density on a logarithmic scale between 0.01 g cm$^{-2}$ and 100 g cm$^{-2}$. The upper right image is at a later time and shows the region within 1 pc of the SMBH with the color scale from 0.025 g cm$^{-2}$ and 250 g cm$^{-2}$. The lower two images are a later time ands show the region within 0.5 pc of the black hole and the colors illustrate the column density on a logarithmic scale between 0.1 g cm$^{-2}$ and 1000 g cm$^{-2}$. Although the cloud was tidally disrupted by the black hole, some of the material is captured by the black hole to form an eccentric disc that quickly fragments to form stars. These are illustrated by the white dots and have eccentricities between e = 0.6 and e = 0.76 and semi-major axes between a = 0.11 pc and a = 0.19 pc. A small population of stars also formed quite early, are visible in the top right panel, and can be seen being ejected from the system in the bottom right panel.

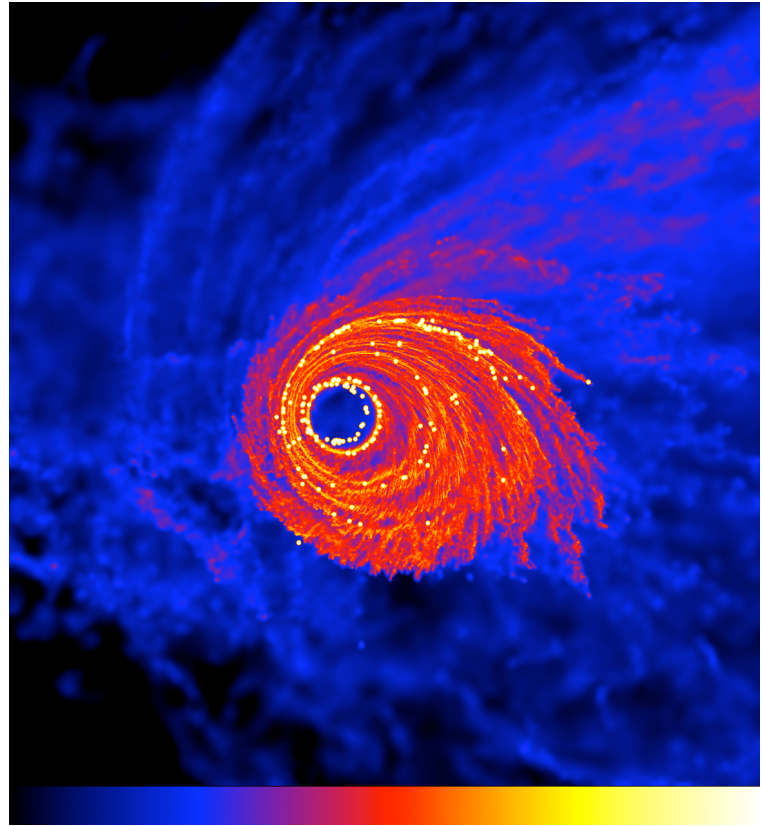

**Fig 2.** The final state of the simulation of a $10^5$ $M_o$ molecular cloud falling towards a $3 \times 10^6$ $M_o$ SMBH. The image shows the region within 0.25 pc of the SMBH located at the centre with the colors illustrating column densities between 0.75 g cm$^{-2}$ and 7500 g cm$^{-2}$. A portion of the cloud has formed a disc around the SMBH, while – at the stage shown here – most of the mass is still outside the region shown. The disc fragments very quickly producing 198 stars with semi-major axes between a = 0.04 pc and a = 0.13 pc and eccentricities between e = 0 and e = 0.53.

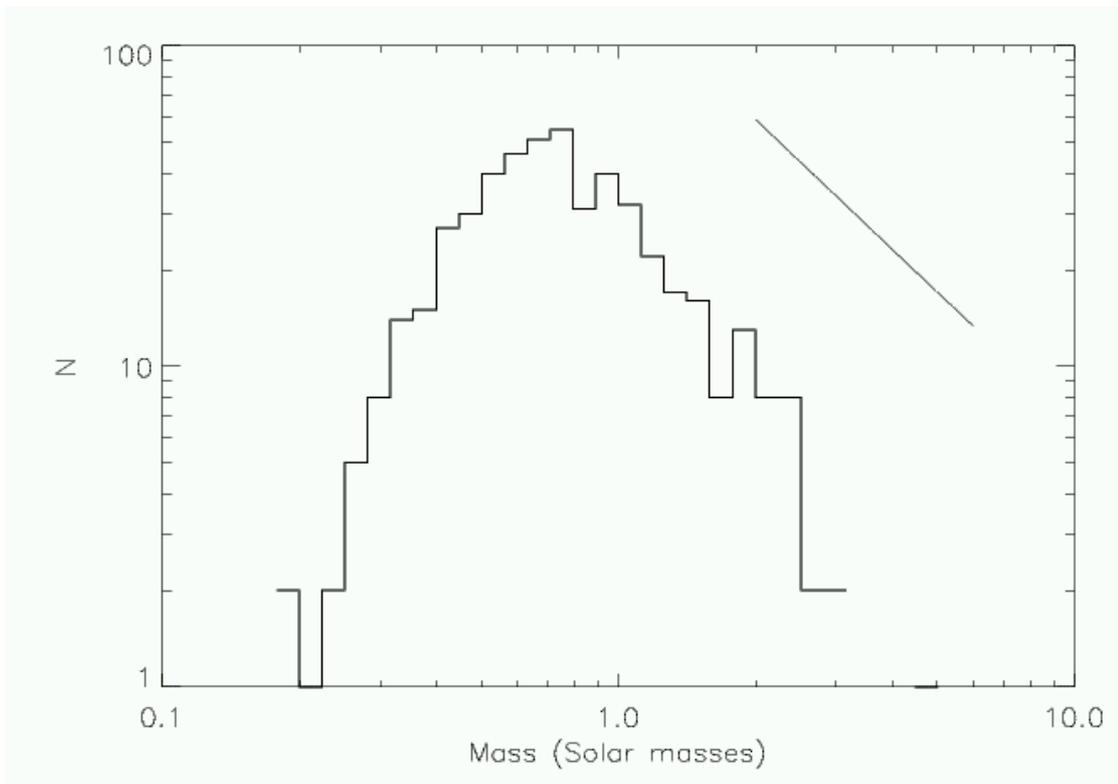

**Fig 3.** Mass function of the stars formed in the simulation illustrated in Fig 1. The stars form with masses close to 0.1 $M_o$, but grow quickly through gas accretion. The mass function there fore has a peak at ~ 0.8 $M_o$, above which it has a power-law form with a slope comparable to that of the salpeter slope, illustrated by the diagonal line.

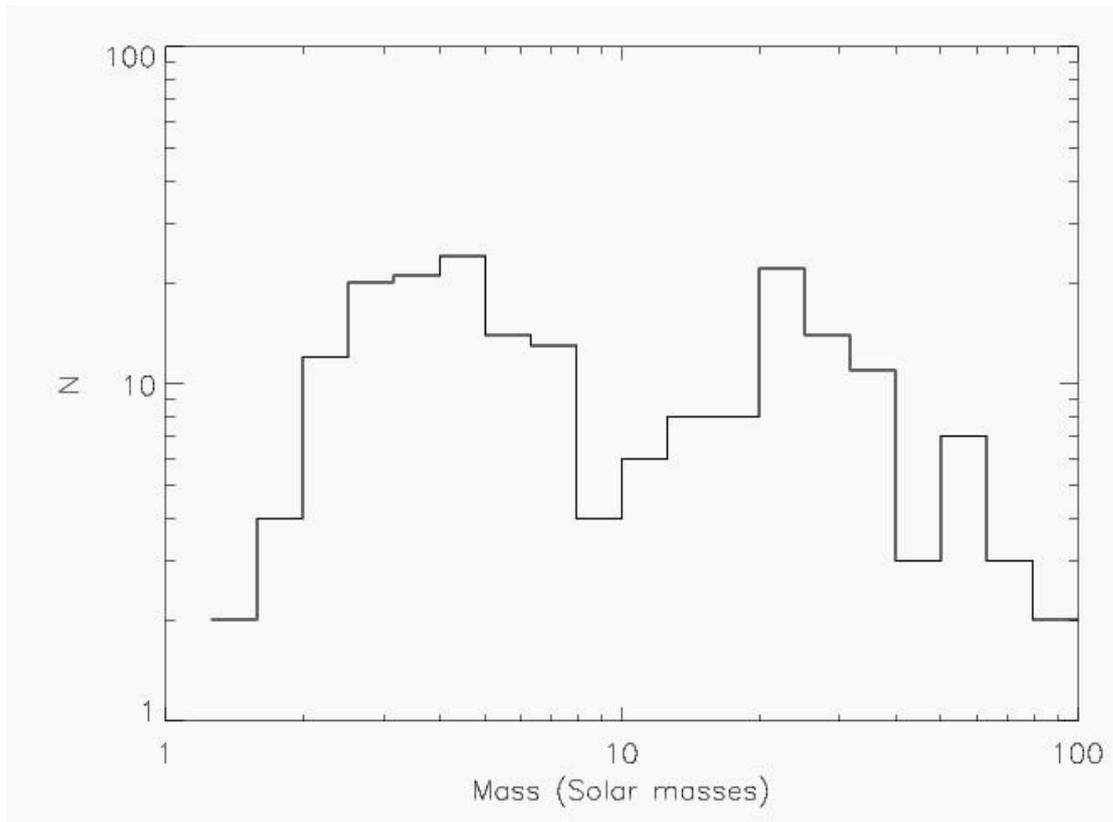

**Fig 4**. Mass function of the stars formed in the simulation illustrated in Fig 2. The mass function is extremely top-heavy and appears to have two populations of stars, a population of massive stars with masses from 10 $M_o$ to 100 $M_o$ and another with masses between 1 $M_o$ and 10 $M_o$.

## Supplemental on-line information

We carried out the simulations using Smooth Particle Hydrodynamics (SPH), a Lagrangian hydrodynamics formalism (17). The two clouds considered are 1) a $10^4$ $M_o$ molecular cloud with a 0.5 pc radius, falling towards a $10^6$ $M_o$ SMBH, and 2) a $10^5$ $M_o$ molecular cloud with a radius of 1 pc falling towards a $3 \times 10^6$ $M_o$ SMBH. The clouds were initially placed 3 pc from the SMBH, infalling at 40 km/s with a tangential velocicty of 8 km/s, yielding an impact parameter of $\sim 0.1$ pc. The clouds were represented by $4.5 \times 10^6$ particles, had an initial temperature of 100 K to account for the background radiation field, and Jeans masses of $\sim 10$ $M_o$. The minimum resolvable protostellar masses were therefore $\sim 0.1$ $M_o$ for the $\sim 10^4$ $M_o$ cloud and 1 $M_o$ for the $10^5$ $M_o$ (20). We also performed a higher-resolution run of the second case where regions of interest had an increased mass resolution down to 0.1 $M_o$., Gas particles that were accreted onto sink-particles (or turned into sink-particles) in the lower resolution run, were split into 9 lower-mass. This mass resolution was always lower than the Jeans mass at the point of fragmentation, ensuring the simulation was adequately resolved. Furthermore, the high-resolution run produced qualitatively and quantitatively similar results. Dense protostellar fragments were replaced by sink-particles (19), which interacted with the rest of the simulation only through gravity and accretion of the gas.

The sink-particles accreted any bound gas particles that fell within 200 AU, while any gas particles that came within 40 AU was accreted, regardless of its properties. Sink-particles were created if the gas particles had a smoothing length smaller than the accretion radius, were bound with a viral parameter <1/2 and collapsing. Creation densities were $\geq 10^{14}$ $M_o$/pc ( $\geq 10^{-8}$ g cm$^{-3}$), well above the critical tidal densities. Only one particle per timestep could be turned into a sink-particle, which immediately accreted all its neighbours ensuring that spurious fragmentation did not occur. The central black hole was modelled by a sink particle with a larger accretion radius of 4000 AU. for bound gas. Encounters between sink-particles had their gravitational forces smoothed by the SPH kernel within 200 AU.

The simulations evolved under a radiative transfer formalism that is described in detail by Stamatellos et al (18). In this formalism, the column density for each particle was determined from its density and gravitational potential energy, the latter reflecting the overall distribution of mass. Rosseland mean opacities (21) were then used to determine the optical depth and consequently the cooling rate, which, together with the hydrodynamic heating rate gave the equilibrium temperature and a thermalisation timescale. The internal energy was updated using an implicit scheme. The equation of state took into account the different chemical states of hydrogen and helium, and the rotational and vibrational degrees of freedom of $H_2$ (22). We assumed that the background radiation field would prevent the cloud from cooling below 100 K. Our computations were performed using the U.K. Astrophysical Fluids Facility (UKAFF) and our local SUPA HPC facility.

The clouds were initially turbulent, with the turbulence modelled by a divergence-free Gaussian velocity field with power spectrum $P(k) \propto k^{-4}$, where $k$ is the wavenumber of the velocity perturbations (23). The velocities were normalized such that the kinetic energy was equal to the absolute magnitude of the potential energy. Including the thermal energy, the clouds were initially marginally unbound but the dissipation of kinetic energy due to shocks ensured the cloud would become globally bound in isolation. Such isolated clouds would dissipate their turbulent energies on the free-fall timescale of $t_{ff} \sim 10^5$ years. Previous simulations of isolated clouds show that the first stars form at 0.5 $t_{ff}$, or 5 x $10^4$ years and continues for 2 $t_{ff}$ (24).

The lower-mass cloud was evolved for 5 x $10^4$ years by which point the cloud had passed the point of closest approach with the black hole. The higher-mass cloud was evolved for 3 x $10^4$ years by which time the infalling gas had formed an eccentric disc that fragmented. The bulk of the latter cloud was still infalling at this point leaving open the possibility of further star formation in this simulation. The largely unbound nature of both infalling clouds, due to the tidal forces from the central black hole, ensured that few stars formed directly from the collapsing cloud. The stars that did form on the way in did not lose significant kinetic energy to become closely bound to the black hole. Instead, they retained their large initial semi-major axis from the black hole. The dissipative nature of the gas was necessary to bind material in close orbits around the black hole. The accretion onto the central black hole in the $10^5$ M$_o$ cloud is also likely to increase from 10, to nearer the 30 % as in the longer evolution of the lower-mass cloud. The unresolved inner disk could also conceivably form additional stars.

The fragmentation of the eccentric disk implies that the Toomre Q-value was Q<1. The infalling material is initially tidally unbound and the Q-value varies between 100 and $10^5$. The rapid increase in the surface density in the eccentric disk forced Q to drop to Q~ 0.1 at the point of fragmentation, in contrast to models where the disk is built up more slowly (10). Surface densities of the disks were of order $10^7$ to $10^8$ $M_o$ $pc^{-2}$. The stars formed at near their Jeans masses (see below) with the majority of their mass being accreted on a timescale of a few thousand years. A total of 6 % of the lower-mass cloud and 3 % of the higher-mass cloud was turned into stars. The average mass accretion rate onto the stars was ~$10^{-4}$ $M_o$ /yr for the lower-mass cloud, $10^{-3}$ $M_o$ /yr for the higher mass cloud with a peak accretion rate of ~ 4 x $10^{-2}$ $M_o$ /yr. These accretion rates would in reality be onto the accretion discs around the forming stars that would modulate the accretion rate over longer time periods. Fragmentation of these disks could conceivably reduce the stellar masses while producing binary systems. It should also be noted that longer-term accretion as the bulk of the higher-mass cloud falls in could also significantly increase the stellar masses.

The resultant stellar masses depend crucially on the gas thermodynamics. Fig 5 plots the temperature distribution of the gas in the higher-mass cloud. In the figure we have plotted every tenth particle plus all particles (red squares) that have densities high enough to prevent them from being tidally disrupted by the supermassive black hole. The compressional heating as the gas fell in and formed the disk dominated over the cooling and increased the temperature to several thousand K. These high temperatures give an initial Jeans mass that vary from several to ~ 50 $M_o$, resulting in high fragment masses and thus a population of massive stars that form quickly in the disk. This is shown in Fig 6 which plots the Jeans mass of every tenth particle plus the

Jeans mass of the neighbour sphere of every particle that has a density high enough to prevent it from being tidally disrupted by the supermassive black hole. The lower mass stars form later once the gas has cooled down to near 100 K. It should be noted, however, that we do not include any feedback from these massive stars and this, if included, could inhibit the later formation of the lower mass stars. This is shown in

The final state of the high-mass simulation is summarised in Fig. 7 which shows the mass distribution of the stars and gas still in orbit around the black hole. The masses of the individual stars and the cumulative mass of the gas is plotted against the virial radius of each sink or gas particle. The highest mass stars are all located in the inner 0.02-0.03 pc with intermediate mass stars found between 0.03 and 0.15 pc. Several stars that formed in the infalling cloud are located at > 1 pc.

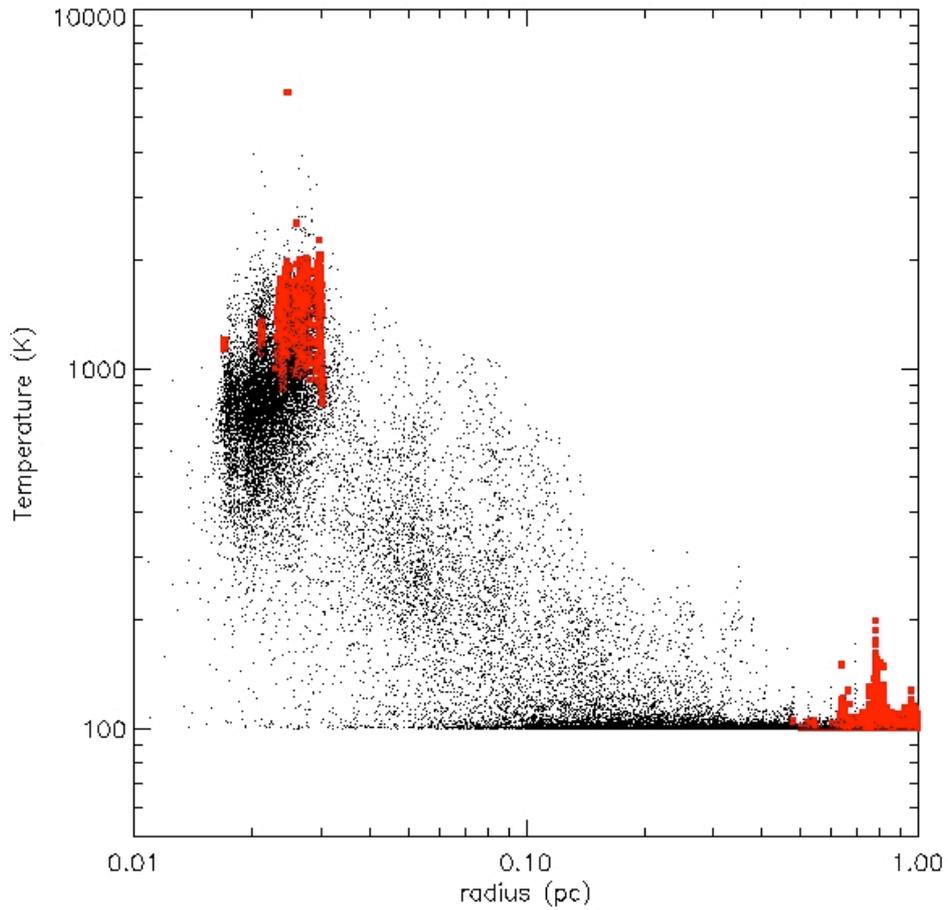

**Fig 5.** Scatter plot showing the particle temperatures plotted against distance from the SMBH. We plot every tenth particle, plus all those particles whose densities are high enough such that they are unlikely to be tidally disrupted by the SMBH. The relatively high temperatures (~ 1000 K) means that the initial Jeans mass is ~ 1 $M_o$ to 50 $M_o$, and those stars that form early are able to grow quickly and become relatively massive.

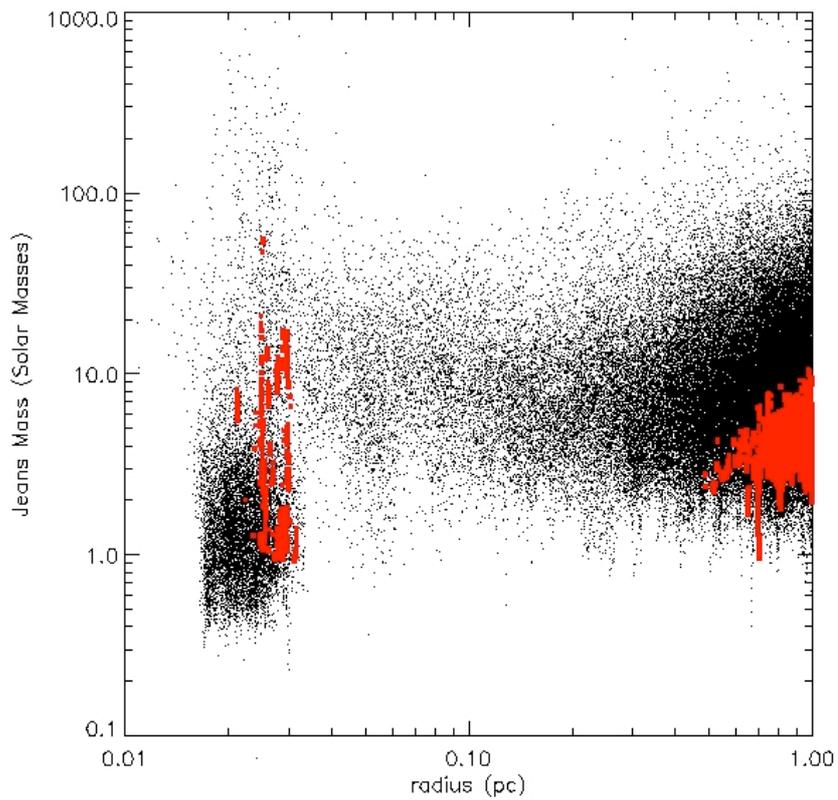

**Fig 6.** Scatter plot showing the Jeans mass plotted against distance from the SMBH. We plot every tenth particle, plus the Jeans mass of the neighbor sphere of all those particles whose densities are high enough such that they are unlikely to be tidally disrupted by the SMBH. The relatively high temperatures (~ 1000 K) means that the initial Jeans mass is ~ 1 $M_o$ to 50 $M_o$, and those stars that form early are able to grow quickly and become relatively massive.

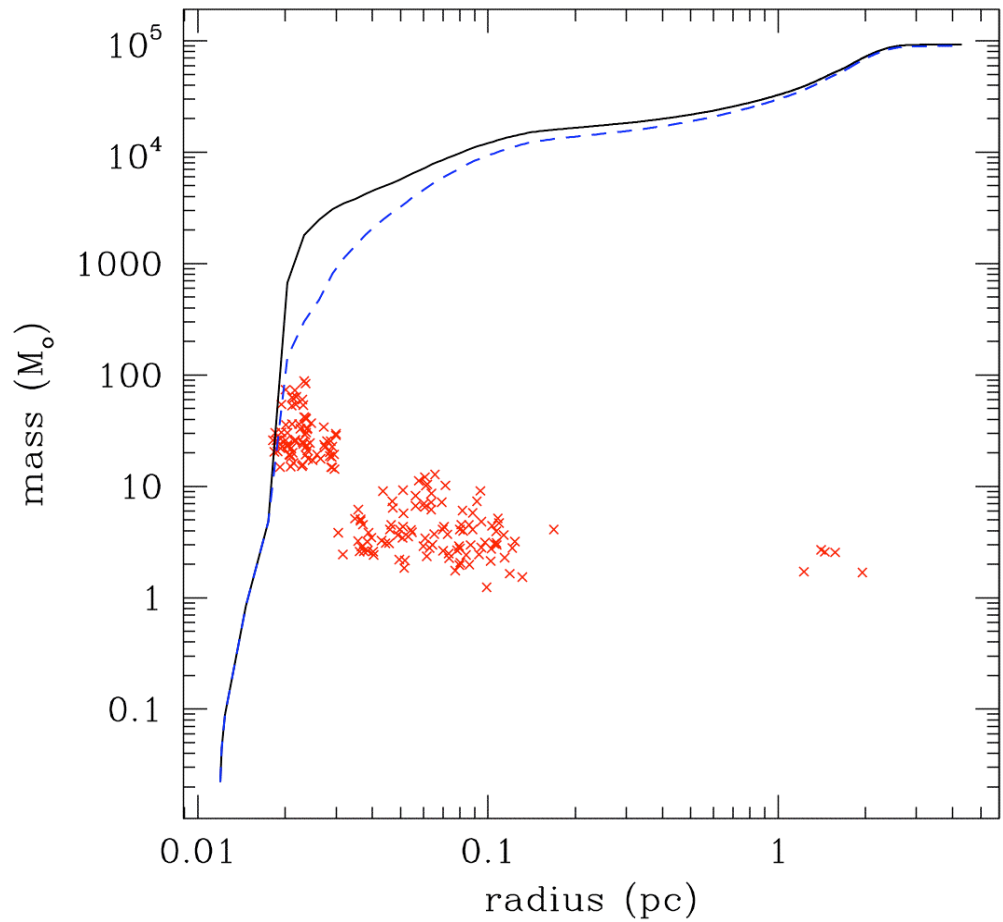

**Fig 7.** The cumulative mass distribution of gas (dashed line) and of gas and stars (solid line) plotted against semi-major axis. The individual stellar masses are also plotted (red crosses) at the location of their individual semi-major axis.